\documentstyle[aps,prd]{revtex}

\def\be{\begin{equation}}
\def\ee{\end{equation}}

\def\z{\chi}
\def\tb{t_{\rm \scriptscriptstyle B}}
\def\Tb{T_{\rm \scriptscriptstyle B}}

\def\Tc{T_{\rm c}}
\def\Ts{T_*}
\def\n#1{n_{\rm #1}}
\def\den#1{\rho_{\rm #1}}
\def\rhol{\rho_{\lambda}}

\begin{document}

\title{Cosmological Phase Transitions in a Brane World}

\author{Stephen C. Davis\thanks{S.C.Davis@swansea.ac.uk},
Warren B. Perkins\thanks{w.perkins@swansea.ac.uk}}
\address{Department of Physics, University of Wales Swansea,\\
Singleton Park, Swansea, SA2 8PP, Wales}

\author{Anne-Christine Davis\thanks{A.C.Davis@damtp.cam.ac.uk} and
Ian R. Vernon\thanks{I.R.Vernon@damtp.cam.ac.uk}}
\address{Department of Applied Mathematics and Theoretical Physics,\\
Centre for Mathematical Sciences,\\
University of Cambridge, Cambridge, CB3 0WA, UK.}

\date{\today}

\maketitle

\begin{abstract}
In brane world scenarios the Friedmann equation is modified, resulting in
an increased expansion at early times. This has important effects 
on cosmological phase transitions which we investigate, 
elucidating significant differences to the standard case. First order
phase transitions require a higher nucleation rate to complete; baryogenesis
and particle abundances could be suppressed. Topological defect evolution
is also affected, though the current defect densities are largely unchanged.
In particular, the increased expansion does not solve the usual monopole
and domain wall problems.
\end{abstract}

\begin{flushright} 
DAMTP-2000-138

SWAT/264B
\end{flushright}

\section{Introduction}

Recently there has been considerable interest in the novel suggestion
that the physical universe is embedded in higher dimensions with
standard model particles confined to a 3-brane and gravity propagating
in the extra dimensions~\cite{brane1,brane2,Sundrum,Rand1}. Randall
and Sundrum~\cite{Rand2} have even suggested that the extra dimension
could be non-compact. Cosmological evolution in these extra dimension scenarios
has been investigated. The Friedmann equation was shown to
contain important deviations from the usual 4-dimensional
case~\cite{Bine1}, giving rise to increased expansion at early times
and a corresponding non-standard temperature-time relation. In 
section~\ref{sec:bwc} we review brane world cosmologies, displaying the
differences to the usual picture.

If the brane world scenario is correct then it needs to describe the
world we live in. In this paper we investigate some of the
implications of the brane world scenario for processes occurring in the
early universe. In particular, the brane cosmology needs to reproduce
standard model physics at low energies and also lead to cosmological
structure formation. Fundamental to both are phase transitions in the
early universe. Since the underlying cosmology is changed in the brane
picture, cosmological phase transitions could also change. In
particular, in section~\ref{sec:pt} we investigate first order phase
transitions in brane cosmologies. The increased expansion at early
times can have dramatic consequences, which we elucidate. We show that
a higher bubble nucleation rate is required for the phase transition
to complete, which could lead to more supercooling in some cases. 

Whilst inflation in brane world scenarios has been
investigated~\cite{wands},  the impact of the increased expansion on
the cosmology of topological defects has yet to be investigated. We
consider the possibility that the five dimensional Planck mass is just
above the grand unified theory (GUT) scale, and consider the effect of
cosmological GUT phase transitions. Such transitions can give rise to
topological defects. We examine whether or not the properties of such
resulting defects are modified in a brane world scenario.  Moving to a
brane world picture introduces  many novel features to defect
cosmology. There are two broad classes of effect that need to be
considered. On the microscopic scale there are changes to the physics
which determines the properties of individual defects. While on the
macroscopic scale, modifications to the Friedmann equation change the
evolution of defect networks. A full treatment of brane world defects
would require a consistent solution representing a defect on the
brane. Given that heavy defects may produce strong gravitational
backreactions on the brane, in general such a calculation lies outside
the scope of the low-energy effective field theory
formalism~\cite{Sundrum}. Some possible changes in defect
microstructure due to the modifications of the gravitational
interaction on small scales are considered in~\cite{inprep}. In this
paper we consider the evolution of {\it standard} defects with the
brane world Friedmann equation in section~\ref{sec:def}. We consider
the effect the increased expansion has on the evolution of cosmic
strings and on the monopole and domain wall densities.

The increased expansion rate in the brane world scenario will lead to
different freeze-out temperatures for particle interactions. This will
change the abundance of particles produced at early times. An example
of this is GUT baryogenesis, which we examine in
section~\ref{sec:bary}. Depending on the fundamental parameters, the
increased expansion rate can lead to a suppression of the resulting
baryon asymmetry. Our conclusions are summarised in
section~\ref{sec:conc}.

\section{Brane World Cosmologies}
\label{sec:bwc}

The recent interest in brane worlds and extra dimensions has
been inspired mostly by~\cite{Rand1}, in which one of the extra 
dimensions is larger than the others and our universe is a positive
tension 3-brane embedded in a five-dimensional bulk. The matter fields
are confined to the brane and following~\cite{Bine2} we consider a
bulk space which contains only a cosmological constant, with energy
density $\rho_B$.

In this case the effective Friedmann equation on the brane is~\cite{ourBW}
\begin{equation} 
\left( \frac{\dot{a}}{a} \right)^2 = 
\left(\frac{4\pi}{3M_5^3}\right) \rho_B  + 
\left(\frac{4\pi}{3M_5^3}\right)^2\rho_b^2 - \frac{k}{a^2} + 
\frac{\mathcal{C}}{a^4} + \frac{F^2}{\rho_b^2 a^8},
\end{equation}
where $M_5$ is the fundamental five-dimensional Planck mass,
$\rho_b$ is the energy density of the brane fields, and $a(t)$ is the
value of the scale factor on the brane. 
$\mathcal{C}$ and $F$ are integration constants; the former is the dark energy
term~\cite{Bine2}, while the latter term is due to a $Z_2$ breaking term in
the metric~\cite{ourBW}. While the energy
density is  $Z_2$ symmetric across the brane, the solution to
Einstein's equations need not be. If we insist on a $Z_2$ symmetric
bulk solution, $F=0$ as in~\cite{Bine2}.

To obtain the standard cosmology at late times we take
$\rho_b = \rho + \rhol$, where $\rhol = 3 M_5^6/(4\pi M_4^2)$ and
$\rho$ are the energy densities corresponding to the brane tension and
matter/radiation on the brane. To obtain a vanishing effective
cosmological constant on the brane the bulk cosmological constant must
be tuned to cancel the brane tension, leading to 
$\rho_B = -4\pi \rhol^2/(3M_5^3)$.
The resulting Friedmann equation is

\begin{equation} \label{Fried_eqtn}
\left( \frac{\dot{a}}{a} \right)^2 = 
\frac{8\pi}{3M_4^2} \rho  + 
\left(\frac{4\pi}{3M_5^3}\right)^2\rho^2 - \frac{k}{a^2} + 
\frac{\mathcal{C}}{a^4} + \frac{F^2}{(\rho + \rhol)^2 a^8}.
\end{equation}

To agree with the standard cosmology the $F$~\cite{ourBW} and
$\mathcal{C}$~\cite{Bine2} terms in Eq.\ (\ref{Fried_eqtn}) must both be
small; this is fully discussed in~\cite{ourBW}. Throughout the rest of this 
paper we will assume they are negligible, and take $k=0$. 

As in the standard case, radiation dominates matter at early times.
Initially in  the brane picture  the $\rho^2$ term dominates Eq.\ 
(\ref{Fried_eqtn}) and we have $a \sim t^{1/4}$.  
After $t=\tb = M_4^2/(8M_5^3)$ the $\rho$ term becomes
dominant and $a \sim (t+\tb)^{1/2}$, as in the standard cosmology.

During the initial period of $\rho^2$ dominance, the
Friedmann equation leads to a non-standard
temperature-time relation,
\be
t = {45 \over 8 \pi^3} {M_5^3 \over g_* T^4}.
\label{branefT}
\ee
This epoch ends when $t=\tb$ and $T=\Tb = [45 M_5^6/(g_* \pi^3
M_4^2)]^{1/4}$.
For $T < \Tb$ the more familiar form is recovered,
\be
t= {3\sqrt{5} \over 4 \pi^{3/2}} {M_4 \over g_*^{1/2} T^2} - \tb. 
\label{branefT2}
\ee
The standard temperature-time relation is Eq.\ (\ref{branefT2}) with $\tb=0$.
As the Hubble parameter in this model is much greater at a given temperature
than in the standard case, the universe initially cools far more rapidly.
This could have dramatic consequences on
cosmological phase transitions, which we investigate in subsequent sections. 

\section{Phase Transitions in Brane Cosmology}
\label{sec:pt}

If the brane world scenario is to describe the Universe we live
in then it must reproduce standard model physics. It must also give rise
to a mechanism for structure formation. In this section we consider 
phase transitions in a brane world; in particular first order phase
transitions. In a cosmological setting, these could be modified because 
of the revised Friedmann equation. 
%We also consider the possibility that
%$M_5$ is just above the GUT scale, and consider cosmological GUT phase
%transitions. These can give rise to topological defects. We examine whether
%or not the properties of such resulting defects are modified in a brane world 
%scenario. 

During a phase transition, bubbles of true vacuum will nucleate and expand. 
Assuming they expand at the speed of light, the fraction of space
remaining in the false vacuum at time $t$ is~\cite{bubbles}
\be
p(t) = \exp \left\{ -{4 \pi \over 3} \int_0^t dt_1 a^3(t_1) \Gamma(t_1) 
	\left[\int_{t_1}^t {dt_2 \over a(t_2)}\right]^3 \right\}.
\label{bubprob}
\ee
The three factors in the integral are respectively the red shift, the
bubble nucleation rate, and the volume at time $t$ of a bubble which
formed at time $t_1$. The probability per unit time and volume that a critical
size bubble will nucleate can be approximated by
$\Gamma(T) = \nu T^4 \theta(\Tc-T)$. The parameter $\nu$ will depend on the
expansion rate~\cite{expbub} as well as the details of the scalar potential, 
thus it will be different in the two models.

The number of bubbles nucleated by time $t$ is given by
\be
\n{bubble} = a^{-3}(t) \int_0^t dt_1 a^3(t_1) \Gamma(t_1) p(t_1).
\label{nbubble}
\ee

\subsection{Standard Cosmology}

Evaluating the integrals in Eq.\ (\ref{bubprob}) in the standard
cosmology gives
\be
p(T) = \exp \left\{ -{25\nu \over 3\pi g_*^2}\left(3 M_4 \over 2\pi\right)^4 
	\left({1\over T} - {1\over\Tc}\right)^4 \right\} \ .
\ee
If the end of the phase transition is taken to be when $p = 1/2$, this
will occur at $T=\Ts$ given by
\be
{1\over\Ts} = {1\over\Tc} + {1.5 g_*^{1/2} \over M_4 \nu^{1/4}} \ .
\label{stdTs}
\ee
For a GUT transition $\Tc \ll M_4$ and so unless $\nu$ is very small
$\Ts \approx \Tc$. 

Evaluating Eq.\ (\ref{nbubble}) reveals that the
number of bubbles nucleated rapidly tends to $0.9 \nu^{3/4} T^3$ after
the end of the transition.

\subsection{Brane Cosmology}

Using Eq.\ (\ref{branefT}) to evaluate Eq.\ (\ref{bubprob}) gives, assuming 
$\Tc > \Tb$,
\be
p(T) = \exp \left\{ -{\pi \nu \over 3}
\left(15 M_5^3 \over 2 \pi^3 g_*\right)^4
	\left({1\over T^3} - {1\over \Tc^3}\right)^4 \right\},
\ee
and so this time the transition ends when
\be
{1\over\Ts^3} = {1\over\Tc^3} + {3.7 g_* \over M_5^3 \nu^{1/4}}.
\label{braneTs}
\ee
If $\Tc$ is close to $M_5$, then even for moderate values of $\nu$ the
second term of Eq.\ (\ref{braneTs}) can be the dominant one. In this case
the transition will be significantly slower and involve a greater
temperature drop. The number of bubbles nucleated, Eq.\ (\ref{nbubble}), has
the same large $t$ behaviour as in the standard cosmology.

\subsection{Avoiding Vacuum Domination}

If the phase transition goes too slowly, it is possible that false
vacuum energy will dominate the radiation before the transition finishes. 
The Universe will then start to inflate, and never stop.

We will consider the simple Higgs model with an effective potential
\be
V(\phi,T) = {\lambda \over 4} \left[ \phi^2 + {T^2 \over 4} 
	- {\Tc^2 \over 4}\right]^2 .
\ee
The false vacuum energy density (when $\phi=0$) is then
\be
\den{FV} = {\lambda \over 64} (\Tc^2 -T^2)^2 p(T).
\ee
To avoid false vacuum domination we need $\den{FV} < \den{radiation}$; hence,
\be
{15 \lambda \over 32 \pi^2 g_*} p(T) < \left(T \over \Tc\right)^4 \ .
\ee

Vacuum domination will only occur if $\nu \ll (\Tc/M_4)^4$, i.e.\ if
the second term of Eq.\ (\ref{stdTs}) is the most significant. Using this
assumption, the bound on $\nu$ for the standard cosmology is
\be
\nu \geq 0.1 g_* \lambda \left(\Tc \over M_4\right)^4 \ .
\label{stdnubound}
\ee
In brane cosmology, assuming $\nu \ll (\Tc/M_5)^{12}$, the
corresponding bound is 
\begin{eqnarray}
\nu &\geq& 4 \times 10^{-3} g_* \lambda^3 \left(\Tc \over
M_5\right)^{12} \nonumber \\ 
&&\approx 8 \times 10^{-3} g_* \lambda \left(\Tc \over M_4\right)^4
\left(\lambda \Tc^4 \over g_* \Tb^4\right)^2 \ .
\end{eqnarray}

Thus unless $\Tc < (g_*/\lambda)^{1/4} \Tb$ phase transitions in the brane
cosmology require a higher nucleation rate to complete successfully.
However, the faster expansion of the universe during the
brane era could allow smaller bubbles to survive, thus increasing the bubble
nucleation rate and weakening the bounds on the underlying
parameters of the theory.
Of course if $\Tc < \Tb$ the phase transition will happen when brane
effects are not significant, and the bound is given by Eq.\ (\ref{stdnubound}).

\section{Defects In Brane Cosmology}
\label{sec:def}

A natural result of cosmological phase transitions are topological defects.
If, after a phase transition, the vacuum manifold has non-trivial
homotopy groups, topological defects will form in brane cosmology, just
as they do in standard cosmology. As in the normal case, defects
can have potentially useful (and sometimes disastrous) cosmological
implications. For example, GUT scale cosmic strings can lead to a 
realistic scenario for structure formation. However, magnetic monopoles
and domain walls rapidly dominate the energy density of the universe.
We examine whether or not the properties of such resulting defects are 
modified in a brane world scenario. To evaluate their properties we need to 
find the initial defect density and then determine how the density evolves.

For a first order phase transition the initial correlation length of
the defects is easily determined from the arguments in the previous section,
\be
\xi \sim {1 \over \lambda \n{bubble}^{1/3}} \sim {1 \over \lambda \Ts} \ .
\ee

Next we consider the evolution of defects in brane cosmology.

\subsection{Shadowing verses Scaling}

An immediate consequence of the modified brany evolution is the 
different relationship between scaling
(i.e.\ a fixed number of defects per horizon volume) and shadowing
(i.e.\ defect density remaining a fixed fraction of the dominant
energy density).

For scaling defects in either model we have
\be
\rho_{\rm string}\propto {ct\over (ct)^3}\propto t^{-2} \ ,
\quad
\rho_{\rm wall}\propto {(ct)^2\over (ct)^3}\propto t^{-1} \ .
\ee

If the dominant energy density varies as $a^{-w}$, we saw above that 
$a\propto t^{1/w}$ in the brane era and $a\propto t^{2/w}$ in the normal
picture. Thus in the brane era we have
\be
\rho_{\rm dominant} \propto t^{-1} \ ,
\ee
while in the normal picture we have
\be
\rho_{\rm dominant} \propto t^{-2} \ .
\ee  
In the standard picture, scaling strings shadow the dominant energy
density, while in the brane era scaling walls shadow the dominant
energy density.

\subsection{Monopoles}

At formation $\n{monopole} \sim \n{bubble}$. Red-shifting gives 
$\n{monopole} \sim T^3$ at later times, thus in the absence of annihilation,
both cosmologies would predict the same monopole number in the current
universe.
While monopole annihilation could look very different in brane cosmology 
if the brane era were persistent, the limited duration of the brane
epoch curtails annihilation.

As for any 2-body annihilation
process, the number density of monopoles relative to photons,
$r_M=n_M/n_\gamma$ is governed by
\be
{d r_M\over dt}= -\beta_M n_\gamma (r_M^2 -r^2_{M_{eq}}) \ ,
\ee
where $\beta_M$ parameterises the monopole annihilation rate and
$r_{M_{eq}}$ is the equilibrium monopole to photon ratio.

Let us assume that $r_{M_{eq}}$ rapidly drops to zero and set
$n_\gamma=\alpha T^3$. We can take the temperature-time relationship
to be  $T=Ct^{-1/w}$, where $C$, $\alpha$ and $w$ are constants.
If the monopoles form at time $t_f$ at a density $r_{M_f}$, integration gives
\be
r_M^{-1}=\beta_M\alpha C^3 {t^{1-3/w}-t_f^{1-3/w}\over
1-3/w} + r_{M_f}^{-1} \ .
\ee
In the standard picture, for a radiation dominated
Universe, $w=2$ and at large times we have the standard 
freeze out picture with
\be
r_M^{-1}\vert_{\rm f.o.} = 
2\beta_M\alpha C^3 t_f^{-1/2} + r_{M_f}^{-1} \ .
\ee
Clearly, freeze out only occurs for $w<3$, for $w>3$, $r_M$ decays  with time.
[For $w=3$, $r_M$ decays like $1/\log(t)$ at late times.]
Thus naively things look 
very different in the brane era. Here $w=4$ and 
\be
r_M^{-1}=4\beta_M\alpha C^3 (t^{1/4}-t_f^{1/4})
+ r_{M_f}^{-1} \ .
\ee
At large times, $r_M\propto t^{-1/4}$, instead of freeze out, the monopole
density continues to decay and there would appear to be no monopole problem.
However, this result does not survive the inclusion of constants and the 
inevitable termination of the brane era.

Let the brane era persist well beyond the GUT time, then
\be
r_M^{-1}\simeq 4\beta_M\alpha C^3 t^{1/4} +r_{M_f}^{-1} \ .
\ee
If we denote the time and temperature of the brane-normal transition
by $\tb$ and $\Tb$,  we have
\be
t_f= \tb \left(\Tb \over T_f\right)^w \ .
\ee
If we now look at the monopole density at $\tb$, in the brane model
we have
\be
r_M^{-1}\simeq 4\beta_M\alpha \Tb^3 \tb +r_{M_f}^{-1} \ .
\ee
While in the normal model we have
\be
r_M^{-1}\vert_{\rm f.o.}
=2\beta_M\alpha \Tb^3 \tb \left(\tb\over t_f \right)^{1/2} +r_{M_f}^{-1} \ .
\ee
Given that $\tb \gg t_f$, we see that the annihilation in the normal
model is far more efficient than in the brane model: the transient
nature of the brane era and the constants conspire to 
over turn the naive expectations from the proportionalites. Thus, the brane 
world scenario does not solve the usual monopole problem associated with 
cosmological phase transitions.

\subsection{Cosmic Strings}

At early times in their evolution cosmic strings
experience a significant damping force from the background radiation
density. For strings the damping force is the dominant effect when
$T^3/\Tc^2 > H$, in which case 
$\xi \propto \Tc T^{-3/2} t^{1/2}$~\cite{defevol}. In the standard
cosmology the correlation length is
\be
\xi \sim {\Tc M_4^{1/2} \over T^{5/2} } \sim {\Tc \over M_4^{3/4}} t^{5/4}, 
\ee
and $\den{string} = \Tc^2 / \xi^2$. This continues until $T \sim
\Tc^2/M_4$, after which the evolution will start to approach a scaling
solution ($\xi \sim t$).

During the initial period of non-standard evolution in the brane
cosmology the string evolution is always friction dominated. This
continues until $T \sim \Tc^2/M_4$, as in the standard cosmology.

For $T>\Tb$ the string density satisfies
\be
\xi \sim {\Tc \over M_5^{3/2} T^{7/2} } \sim {\Tc \over M_5^{9/8}} t^{7/8}.
\ee
Using the relationship between $\Tb$, $M_5$ and $M_4$, we find that
the correlation length at $T=\Tb$ in the brane cosmology is of the
same order as in the standard cosmology. For $T<\Tb$ the correlation
length  evolves as in
the standard case, thus the string density at the end of friction
domination is the same in both pictures.

\subsection{Domain Walls}

The early evolution of domain walls can also be friction
dominated. The correlation length is $\xi \sim vt$, where $v$ is the
speed of the walls. During friction domination the wall tension and
friction will of the same order. This determines the speed of the
walls: $v^2 \sim \Tc^3/(t T^4)$~\cite{defevol}.
Now $\den{wall} \sim \Tc^3/\xi$ so 
$\den{wall} / \den{rad.} \sim v$. This means that when the walls
become relativistic they will also start to dominate the energy
density of the Universe.

In the standard cosmology the velocity of the walls is
\be
v \sim {\den{wall} \over \den{rad.}} \sim {\Tc^{3/2} \over M_4^{1/2} T} \ .
\ee
Hence the walls dominate the Universe at $T= (\Tc/M_4)^{1/2} \Tc$ \ .

During the early brane cosmology,
\be
v \sim {\den{wall} \over \den{rad.}} 
  \sim \left(\Tc \over M_5 \right)^{3/2} = \mbox{const.}
\ee
Thus the domain wall energy density initially scales like
radiation. After $T=\Tb$ they will behave as in the standard picture. As in
the case with monopoles, the brane world scenario does not stop domain
walls dominating the energy density of the universe, despite naive
expectations.

\section{GUT Baryogenesis}
\label{sec:bary}

In this section we investigate the modifications of the usual
picture of baryogenesis which result from the brane world Friedmann equation. 
In general there are three things which are required for successful
baryogenesis~\cite{Sakarov}:  (1) baryon number violation, (2) $C$ and
$CP$ violation and (3) departure from thermal equilibrium. The third
requirement can be illustrated with a simple generic model in which
the baryon asymmetry is produced by the decay of GUT bosons ($X$,
$\bar X$)~\cite{bary}. At high temperatures ($T \gtrsim m_X$) the
$X$-bosons behave relativistically and so  $n_X = n_{\bar X} \simeq
n_\gamma$  in equilibrium. If the $X$-bosons are still in thermal
equilibrium for $T \lesssim m_X$, $n_X = n_{\bar X} \simeq
(m_X/T)^{3/2} \exp (-m_X/T) \ll n_\gamma$, and so when they eventually
decay they will produce exponentially few baryons. On the other hand
if the $X$-bosons decouple before $T \sim m_X$ there will be
$n_\gamma$ of them to decay into baryons. In terms of the possible
annihilation and decay processes, baryogenesis arises from the single
particle decay of $X$ and $\bar X$'s rather than $X\bar X$
annihilation.

The interactions which determine the effectiveness of the above
mechanism are the decays and inverse decays of $X$-bosons, and 
$2 \leftrightarrow 2$ $X$-mediated $B$-nonconserving scatterings between
baryons. For $T \lesssim m_X$ rates of these processes are, respectively,
\be
\Gamma_D \simeq \alpha m_X,
\ee
\be
\Gamma_{ID} \simeq \alpha m_X
	\left(m_X \over T\right)^{3/2}  \exp \left(-m_X \over T\right),
\ee
and
\be
\Gamma_{BNC} \simeq A\alpha^2 {T^5 \over m_X^4},
\ee
where $\alpha$ measures the coupling strength of the $X$-boson and
$A$ is a large numerical factor which accounts for the number of
scattering channels. If the inverse decays or the $2 \leftrightarrow 2$ 
baryon scatterings are still significant for $T \lesssim m_X$ the
final baryon asymmetry will be suppressed. 

It is convenient to define the parameters $\z = m_X/T$ and $K =
\left. \Gamma_D / H \right|_{\z=1}$. In the standard cosmology,
\be
K \simeq {\alpha M_4 \over g_*^{1/2} m_X} \ ,
\ee
while in the brane cosmology,
\be
K \simeq {\alpha M_5^3 \over g_* m_X^3} \ .
\ee
The ratios of the interaction rates to the Hubble parameter are then
\be
\Gamma_{ID}/H \simeq K \z^{w+3/2} e^{-\z} \ ,
\ee
\be
\Gamma_{BNC}/H \simeq K A\alpha \z^{w-5} \ ,
\ee
where $w=2$ for the standard cosmology and $w=4$ for the brane
cosmology.

If $K \ll 1$ then $\Gamma_{ID}/H < 1$ and $\Gamma_{BNC}/H < 1$ at
$\z=1$, thus the $X$-bosons decouple when they are relativistic and the
baryon asymmetry ($B$) will be maximal. If each $X$ decay produces a
mean net baryon number $\epsilon$, then the final baryon number to
entropy ratio produced when $K \ll 1$ will be
\be
B = {n_b - n_{\bar b} \over g_* n_\gamma} \simeq {\epsilon \over g_*} \ .
\label{Bmax}
\ee
The $C$ and $CP$ violation parameter $\epsilon$ is of order $\alpha^N$,
where $N \geq 1$ since this is not a tree level process.

This baryon asymmetry can be damped if either the inverse decays or the
baryon non-conserving scatterings persist beyond $\z=1$. 
If $1 \lesssim K \lesssim K_C$ (where $K_C$ is a
theory dependent constant), the inverse decays will still be
significant for $\z > 1$. This continues until the inverse decays
freeze out with $\Gamma_{ID}/H \simeq \z$ at $\z=\z_f$. 
Approximate integration of the Boltzmann equation in this case gives
\be
B \simeq {\epsilon \over g_* K \z_f^{w-1}} \ .
\label{Bsup}
\ee
For large $K$, $\z_f$ has a slow, logarithmic dependence on $K$ and the baryon
asymmetry falls roughly as the inverse of $K$. 

If $K \gtrsim K_C$ the $B$-nonconserving scatterings will provide the
dominant damping mechanism. The freeze out for these interactions
is determined by $\Gamma_{BNC}/H \simeq \z$. In this case, 

\be
B \simeq {\epsilon \over g_*} \z_f^2 \exp\left(-\frac{6-w}{5-w}\z_f\right) \ .
\label{Bexp}
\ee
Thus for large $K$ the baryon asymmetry is exponentially suppressed as
expected. $K_C$ is determined by the value of $K$ that gives simultaneous
freeze out of both the inverse decays and the baryon non-conserving 
scatterings.

To compare the two cases we will consider typical GUT parameters:
$g_* = 200$, $A = 5000$, gauge coupling strength $\alpha_G = 1/45$ 
and Higgs coupling strength  $\alpha_H = 10^{-3}$.

In the standard cosmology these parameters give
\be
K\sim\cases{10^{-3} M_4/m_X & gauge \cr
           10^{-4} M_4/m_X & Higgs \cr},
\quad
K_C\sim\cases{120  & gauge \cr
             15000 & Higgs \cr}.
\ee

Assuming $m_X \gtrsim 10^{14}\,$GeV, there is no damping of $B$ in the
Higgs boson mediated case, while in the gauge boson mediated case $B$
is power law damped. The $CP$ violation required to obtain
the observed baryon density, $B \sim 10^{-10}$, is given by
\be
\epsilon\sim \cases{10^{-7} (10^{16}\mbox{GeV}/m_X) & gauge\cr
                    10^{-8} & Higgs}.
\ee

The corresponding values of $K$ and $K_C$ in the brane cosmology are
\be
K \sim \cases{ (M_5/20m_X)^3 & gauge \cr
               (M_5/60m_X)^3 & Higgs \cr}
,\quad
K_C \sim \cases{ 1 & gauge \cr
               60 & Higgs \cr}
.
\ee

Thus in the gauge boson mediated case, unless $m_X$ is within an order
of magnitude  of $M_5$, the baryon asymmetry will be exponentially
suppressed. In the Higgs boson case $m_X$ must be within two orders of
magnitude of $M_5$ to give significant baryogenesis. This change
renders GUT  baryogenesis particularly sensitive to an early brane
era. If $m_X\sim 10^{14}\,$GeV, baryogenesis occurs in the brane era
for $M_5$ as high as $10^{16}\,$GeV.

However, this result was obtained using typical standard cosmology GUT
values. 
If $M_5$ were far lower than the usual Planck scale, then the GUT
scale would also have to be reduced. $\alpha$ would also have to be
substantially smaller in order to avoid breaking the experimental
bounds on the proton lifetime. This would result in a lower
$\epsilon$, further constraining the model.

\section{Conclusions} 
\label{sec:conc}

In this paper we have considered the implications of brane world 
models for various processes which occur in the early universe.
Due to the modified Friedmann equation, the rate of expansion of the 
Universe is increased at early times. The relation between
temperature and time is also changed. This has important phenomenological
consequences. 

As discussed in section~\ref{sec:pt}, first order phase
transitions require a higher nucleation rate in order to complete, which
could result in more supercooling. Indeed,
if the nucleation rate is not high enough, the Universe becomes dominated
by the false vacuum and the transition does not complete.

Processes that rely on interactions freezing out are also sensitive to the
enhanced expansion rate. For example, this has important consequences for 
GUT baryogenesis. As shown in section~\ref{sec:bary}, unless the mass of 
the relevant GUT particle is within two orders of magnitude of the fundamental 
Planck scale, the baryon excess is exponentially suppressed. More generally,
the abundance of any species that freezes out during the brane epoch
will be affected.

Defect evolution is also modified during the brane epoch. However, due to
the transient nature of this phase, the current defect densities are
largely unchanged. This suggests that the usual mechanism for defect 
inspired structure formation is largely unchanged. It also suggests that,
despite the increased expansion rate at early times, the usual monopole 
problem associated with GUT models remains.  

\acknowledgements
We would like to thank Pierre Binetruy, Tim Hollowood, Nathalie Deruelle,
and David Wands for discussions and PPARC for financial support.

\def\Journal#1#2#3#4{{#1}{\bf #2}, #3 (#4)}
\def\npb{Nucl.\ Phys.\ {\bf B}}
\def\jpa{J.\ Phys.\ {\bf A}}

\end{document}